\documentclass{ifacconf}

\usepackage{graphicx}      
\usepackage{natbib}        
\usepackage{amsmath} 
\usepackage{amssymb}  
\usepackage{mathrsfs}
\usepackage{color}
\usepackage{bm}
\usepackage{bbm}
\usepackage{subfigure}
\usepackage{supertabular}

\newcommand{\be}[1]{\begin{equation}\label{#1}}
\newcommand{\benon}{\begin{equation*}}  
\newcommand{\bemuln}[1]{\begin{multline}\label{#1}}
\newcommand{\bemul}{\begin{multline*}}
\newcommand{\bee}{\begin{eqnarray*}}
\newcommand{\eee}{\end{eqnarray*}}
\newcommand{\been}[1]{\begin{eqnarray}\label{#1}}
\newcommand{\eeen}{\end{eqnarray}}
\newcommand{\began}[1]{\begin{gather}\label{#1}}
\newcommand{\bega}{\begin{gather*}}
\newcommand{\bealn}[1]{\begin{align}\label{#1}}
\newcommand{\beal}{\begin{align*}}
\newcommand{\bealatn}[2]{\begin{alignat}{#1}\label{#2}}
\newcommand{\bealat}{\begin{alignat*}}
\newcommand{\bexalatn}[1]{\begin{xalignat}\label{#1}}
\newcommand{\bexalat}{\begin{xalignat*}}

\newcommand{\mb}{\mathbf}

{\theoremstyle{break} \theorembodyfont{\it}
\newtheorem{defi}{Definition}
\newtheorem{ass}{Assumption} 
\newtheorem{rmk}{Remark} 
 }

\def\bb{{\mathbf b}}

\def\bd{{\mathbf d}}

\def\bg{{\mathbf g}}
\def\bh{{\mathbf h}}

\def\bm{{\mathbf m}}

\def\bt{{\mathbf t}}

\def\bw{{\mathbf w}}
\def\bx{{\mathbf x}}  
\def\by{{\mathbf y}}
\def\bz{{\mathbf z}}
\def\bA{{\mathbf A}}

\def\bN{{\mathbf N}}

\def\bP{{\mathbf P}}
\def\bQ{{\mathbf Q}}

\def\bS{{\mathbf S}}

\def\texitem#1{\par\smallskip\noindent\hangindent 25pt
               \hbox to 25pt {\hss #1 ~}\ignorespaces}

\newcommand{\scrA}{\mathcal{A}}

\newcommand{\scrF}{\mathcal{F}}

\newcommand{\scrI}{\mathcal{I}}

\newcommand{\scrR}{\mathcal{R}}
\newcommand{\scrS}{\mathcal{S}}

\newcommand{\scrW}{\mathcal{W}}

\newcommand{\bbeta}{\boldsymbol{\beta}}

\newcommand{\bepsilon}{\boldsymbol{\epsilon}}

\newcommand{\bPhi}{\boldsymbol{\Phi}}
\newcommand{\bxi}{\boldsymbol{\xi}}

\DeclareMathOperator{\VI}{\text{VI}}

\usepackage{algorithm}
\usepackage{algorithmicx,algpseudocode}

\makeatletter
\newcommand*{\defeq}{\stackrel{\text{def}}{=}}
\makeatother

\makeatletter
\newcommand{\rmnum}[1]{\romannumeral #1}
\newcommand{\Rmnum}[1]{\expandafter\@slowromancap\romannumeral #1@}
\makeatother

\begin{document}
\begin{frontmatter}

\title{Data-driven Estimation of Origin-Destination Demand and User Cost Functions for the Optimization of Transportation Networks\thanksref{footnoteinfo}} 

\thanks[footnoteinfo]{Research partially supported by the
	NSF under grants CNS-1239021, ECCS-1509084, CCF-1527292,
	IIS-1237022, and IIP-1430145, by the AFOSR under grant
	FA9550-15-1-0471, by the ARO under grants W911NF-11-1-0227 and
	W911NF-12-1-0390, and by The MathWorks.}

\author[First]{Jing Zhang} 
\author[First]{Sepideh Pourazarm} 
\author[Second]{Christos G. Cassandras}
\author[Second]{Ioannis Ch. Paschalidis}

\address[First]{Division of Systems Engineering, Boston University\\ (e-mail: \{jzh, sepid\}@bu.edu)}
\address[Second]{ Department of Electrical and
	Computer Engineering and Division of Systems Engineering, Boston University\\ (e-mail: \{cgc, yannisp\}@bu.edu)}

\begin{abstract}                

In earlier work (Zhang et al., 2016) we used actual traffic data from the Eastern Massachusetts transportation network in the form of spatial average speeds and road segment flow capacities in order to estimate Origin-Destination (OD) flow demand matrices for the network. Based on a Traffic Assignment Problem (TAP) formulation (termed ``forward problem''), in this paper we use a scheme similar to our earlier work to estimate initial OD demand matrices and then propose a new inverse problem formulation in order to estimate user cost functions. This new formulation allows us to efficiently overcome numerical difficulties that limited our prior work to relatively small subnetworks and, assuming the travel latency cost functions are available, to adjust the values of the OD demands accordingly so that the flow observations are as close as possible to the solutions of the forward problem. We also derive sensitivity analysis results for the total user latency cost with respect to important parameters such as road capacities and minimum travel times. Finally, using the same actual traffic data from the Eastern Massachusetts transportation network, we quantify the Price of Anarchy (POA) for a much larger network than that in Zhang et al. (2016).

\end{abstract}


\end{frontmatter}

\section{Introduction} \label{sec:intro}

Using actual traffic data from a transportation network, we measure its performance by using the total travel time all drivers experience across different Origin-Destination (OD) pairs. Our objective is to compare two different strategies: a user-centric strategy vs. a system-centric one and quantify the difference between the two through the Price of Anarchy (POA) metric.

More specifically, for a user-optimal strategy we face a system with non-cooperative agents (drivers) in which each individual driver follows her personally optimal policy without caring about the system performance. Such ``selfish behavior'' results in a Nash (Wardrop) equilibrium, a point where no agent can benefit by altering its actions assuming that the actions of all the other agents remain fixed (see \cite{Youn2008}). The equilibrium flows in traffic networks, known as the ``Wardrop equilibrium," (see \cite{patriksson1994traffic}) is the solution to the Traffic Assignment Problem (TAP) that will be described in the next section. It is known that the Nash equilibrium is usually suboptimal from the system point of view. 

On the other hand, under a system-optimal policy we enjoy a system in which all agents (drivers) cooperate in order to achieve a social optimum, which is the best policy for the society as a whole (see \cite{Youn2008}). 
To assess the scope of improvement in a transportation system, we define the Price of Anarchy (POA) as a measure to quantify the system inefficiency under a user-optimal strategy vs. a system-optimal strategy. The latter can be implemented, for example, by taking advantage of the emergence of Connected Automated Vehicles (CAVs) through which socially optimal routing decisions can be executed.

In \cite{CDC16} we have studied a similar problem using actual traffic data from the Eastern Massachusetts (EMA) road network. In particular we have investigated the system performance of a relatively small interstate highway subnetwork of EMA. 

The first goal of this paper is to quantify the POA  for a general single-class transportation network model. We take similar steps as in \cite{CDC16} which include inferring equilibrium flows from the speed dataset, estimating OD demand matrices, estimating per-road travel latency cost functions, and calculating social optimum flows in order to quantify the POA. In contrast to \cite{CDC16}, here we are able to analyze a much larger subnetwork of the EMA road network in the sense that it includes a much larger number of nodes and links.

To achieve this goal, we first propose an alternative formulation for the travel latency cost function estimation problem, which enables us to overcome numerical difficulties when the network gets larger. We then modify existing schemes to perform OD demand estimation for large-scale road networks, regardless of whether they exhibit congestion or not. 
 In particular, we leverage a bi-level optimization problem formulation allowing us to estimate OD demand matrices for a given larger network based on the OD demands of its representative (landmark) subnetworks. 
 Utilizing the estimated travel latency cost functions and OD demand matrices, we then quantify the POA.

Another goal of this paper is to analyze the sensitivities of the total users' travel latency costs with respect to key parameters such as road capacities and minimum travel times. The results would help prioritize road segments for interventions that can mitigate congestion. We derive sensitivity analysis formulae for the TAP and apply them to the same interstate highway subnetwork of EMA as that in \cite{CDC16}.

The rest of the paper is organized as follows: In Sec. \ref{cdc16-sec:mod} we introduce models and methods that we use. In Sec. \ref{sec: dataset} we describe the datasets and explain the data processing procedures. Numerical results are shown in Sec. \ref{Sec:Rsults}. 

\textbf{Notation:} All vectors are column vectors. For economy of space, we write $\bx = \left(x_1, \ldots, x_{\text{dim}(\bx)}\right)$ to denote the column vector $\bx$, where $\text{dim}(\bx)$ is the dimension of $\bx$. We use prime ($\prime$) to denote the transpose of a matrix or vector. Unless otherwise specified, $\|\cdot\|$ denotes the
$\ell_2$ norm. $\left| \mathcal{D} \right|$ denotes the cardinality of a
set $\mathcal{D}$. $A \defeq B$ indicates $A$ is defined using
$B$.

\section{Models and Methods} \label{cdc16-sec:mod}

\subsection{Model for a single-class transportation network}

We begin by reviewing the model of \cite{CDC16}.
Denote a road network by $\left( {\mathcal{V}, \scrA, \mathcal{W}} \right)$, where $\left( {\mathcal{V}, \scrA}
\right)$ forms a directed graph with $\mathcal{V}$ being the set of nodes and $\scrA$ the
set of links, and $\mathcal{W} = \left\{ {{\textbf{w}_i}:{\textbf{w}_i} = \left(
	{{w_{si}},{w_{ti}}} \right), \,i = 1, \ldots ,\left| \mathcal{W} \right|}
\right\}$ indicates the set of all OD pairs. Assume the graph $\left( {\mathcal{V}, \scrA}
\right)$ is strongly connected and let $\textbf{N} \in {\left\{
	{0,1, - 1} \right\}^{\left| \mathcal{V} \right| \times \left| \scrA
		\right|}}$ be its node-link incidence matrix. Denote by $\textbf{e}_{a}$ the
vector with an entry being 1 corresponding to link $a$ and all the other
entries being 0.
For any OD pair $\textbf{w} = \left( {{w_s},{w_t}} \right)$, denote by ${d^{\textbf{w}}} \ge 0$ the
amount of the flow demand from $w_s$ to $w_t$. Let ${\textbf{d}^{\textbf{w}}}
\in {\mathbb{R}^{\left| \mathcal{V} \right|}}$ be the vector which is all
zeros, except for a $-{d^{\textbf{w}}}$ (resp., ${d^{\textbf{w}}}$) in the coordinate mapping to node
$w_s$ (resp., $w_t$).

Denote by $\mathcal{R}_{i}$ the set of simple routes (a route without cycles is called ``\textit{simple route}'') between OD pair $i$.  For each $a \in \scrA$, $i \in \{1, \ldots ,\left| \mathcal{W} \right|\}$, $r \in
\mathcal{R}_{i}$, define the link-route
incidence by
\[
{\delta ^{i}_{ra}} = \left\{ \begin{gathered}
1,{\text{  if route }}r \in {\mathcal{R}_{i}}{\text{ uses link }}a, \hfill \\
0,{\text{  otherwise.}} \hfill \\
\end{gathered}  \right.
\]

Let $x_a$ denote the flow on link $a \in \scrA$ and $\bx \defeq (x_a; \; a \in \scrA)$ the flow vector.  Denote by ${t_a(\bx)}:\mathbb{R}_ + ^{\left|
	\scrA \right|} \to {\mathbb{R}_ + }$ the \textit{travel latency cost} (i.e., \textit{travel time}) function for link $a
\in \scrA$. If for all $a \in \scrA$, ${t_a(\bx)}$ only depends on
$x_a$, we say the cost function $\bt\left( \bx \right) = \left( {{t_a}\left( {{x_a}} \right); \; a \in \scrA} \right)$ is \textit{separable} (see \cite{patriksson1994traffic}).
Throughout the paper, we assume that the travel latency cost functions are separable and take the following
form (see \cite{bertsimas2014data} \& \cite{branston1976link}):
\begin{align}
{t_a}\left( {{x_a}} \right) = {t^{0}_a}f\left( {\frac{{{x_a}}}{{{m_a}}}} \right), \label{cdc16-costf}
\end{align}
where $t^{0}_a$ is the \textit{free-flow travel time} of $a \in
\scrA$, $f(0)=1$, $f(\cdot)$ is strictly increasing and continuously
differentiable on $\mathbb{R}_+$, and $m_a$ is the \textit{flow capacity} of
$a \in \scrA$.

Define the set of feasible flow vectors $\scrF$ as (see \cite{bertsimas2014data})
\[
\left\{ {\bx:\exists {\bx^{\textbf{w}}} \in \mathbb{R}_ +
	^{\left| \scrA \right|} ~\text{s.t.}~\bx =
	\sum\limits_{\textbf{w} \in \mathcal{W}} {{\bx^{\textbf{w}}}}
	,\textbf{N}{\bx^{\textbf{w}}} = {\textbf{d}^{\textbf{w}}},\forall
	\textbf{w} \in \mathcal{W}} \right\},
\]
where $\bx^\bw$ indicates the flow vector attributed to OD pair $\bw$.
In order to formulate appropriate optimization and inverse optimization problems arising in transportation networks, we state
the definition of \textit{Wardrop equilibrium}:

\begin{defi} \label{cdc16-def1} \em{(\cite{patriksson1994traffic}).}  A feasible flow $\bx^* \in \scrF$ is a {\em{Wardrop equilibrium}} if for every OD pair
	$\bw \in \mathcal{W}$, and any route connecting $(w_s,w_t)$ with
	positive flow in $\bx^*$, the cost of traveling along that route is
	no greater than the cost of traveling along any other route that
	connects $(w_s,w_t)$. Here, the cost of traveling along a route is the sum of
	the costs of each of its constituent links.
\end{defi}

\subsection{The forward problem}  \label{sec:forward}

As in \cite{CDC16}, here we refer to the Traffic Assignment Problem (TAP) as the \textit{forward problem}, whose goal is to find the
Wardrop equilibrium for a given single-class transportation network with a
given OD demand matrix.  
It is a well-known fact that, for network $\left( {\mathcal{V}, \scrA, \mathcal{W}} \right)$, the TAP can be formulated as the following
optimization problem (see \cite{dafermos1969traffic} \& \cite{patriksson1994traffic}):
\begin{align}
\text{(TAP)} \quad \mathop {\min }\limits_{{\bx \in \scrF}} \sum\limits_{a \in \scrA} {\int_0^{{x_a}} {t_a(s)ds} }. \label{cdc16-tap}
\end{align}
As an alternative, we also formulate the TAP as a Variational
Inequality (VI) problem:

\begin{defi}
\label{cdc16-def2} \em{(\cite{bertsimas2014data})}. The VI problem, denoted as $\text{VI}\left( {\bt,\scrF} \right)$, is to find an ${\bx^ * } \in \scrF$ s.t.
\begin{align}
\bt{\left( {{\bx^ * }} \right)'}\left( {\bx - {\bx^ * }} \right) \geq 0, \quad \forall \bx \in \scrF. \label{cdc16-VI}
\end{align}
\end{defi}

We introduce an assumption:

\begin{ass}
\em{}
\label{cdc16-assumption1}
$\bt(\cdot)$ is strongly monotone (see \cite{patriksson1994traffic} or
\cite{bertsimas2014data} for the definition of strong monotonicity) and
continuously differentiable on $\mathbb{R}_ + ^{\left| \scrA
	\right|}$. $\scrF$ is nonempty and contains an interior point
(Slater's condition).
\end{ass}

\begin{thm} \label{cdc16-th21} (\cite{patriksson1994traffic}).
 \textit{Assump. \ref{cdc16-assumption1}} implies that there exists
	a Wardrop equilibrium of the network $\left( {\mathcal{V}, \scrA, \mathcal{W}} \right)$, which is the unique solution to $\text{VI}(\bt,\scrF)$.
\end{thm}

\subsection{The inverse problem} \label{cdc16-sec InverseVI-uni}

To solve the forward problem, we need to know the travel latency cost function and the OD demand matrix. Assuming that we know the OD demand matrix and have observed the Wardrop equilibrium flows, we seek to formulate the inverse problem (the inverse VI problem, in particular), so as to estimate the travel latency cost function.

For a given $\epsilon > 0$, we define an
\textit{$\epsilon$-approximate solution} to $\VI(\mathbf{t}, \scrF)$ by
changing the right-hand side of \eqref{cdc16-VI} to $- \epsilon$:

\begin{defi}
\label{cdc16-def3}
\em{(\cite{bertsimas2014data})}. Given $\epsilon > 0$,
$\hat{\bx} \in \scrF$ is called an
\textit{$\epsilon$-approximate solution} to $\VI(\mathbf{t}, \scrF)$ if
\begin{equation}
\label{cdc16-eq:defApproxEquil}
\mathbf{t}(\hat{\bx})'(\bx - \hat{\bx}) \geq - \epsilon, \quad
\forall \bx \in \scrF.
\end{equation}
\end{defi}

Given $K$ observations $(\bx_k, \scrF_k)$, $k = 1, \ldots, K$,
with $\bx_k \in \scrF_k$ and each $\scrF_k$ being a set of feasible flow
vectors meeting Slater's condition (see \cite{boyd2004convex}), the inverse VI problem
amounts to finding a function $\bt$ such that $\bx_k$ is an
$\epsilon_k$-approximate solution to $\VI(\bt, \scrF_k)$ for each $k$.
Denoting $\bepsilon \defeq (\epsilon_k; \; k = 1, \ldots, K)$, we can formulate the inverse VI problem as (see \cite{bertsimas2014data})
\begin{align}
\min_{{\mathbf{t}}, \bepsilon} \quad & \| \bepsilon \|
\label{cdc16-inverseVI1} \\
\text{s.t.} \quad & {\mb{t}}({\bx_k})'(\bx-{\bx_k}) \geq -\epsilon_k, \quad \forall
\bx\in \scrF_k,\ \forall k, \notag \\
& \epsilon_k > 0, \quad \forall k. \notag
\end{align}

To make \eqref{cdc16-inverseVI1} solvable, we apply the nonparametric estimation approach
which expresses the cost function in a Reproducing Kernel Hilbert Space
(RKHS) (see \cite{bertsimas2014data} \& \cite{evgeniou2000regularization}). To be specific, we pick the polynomial
kernel, i.e., ${\phi}(x, y) \defeq ( c + xy )^n$ for some choice of
$c \geq 0$ and $n \in \mathbb{N}$ (for the specifications of $c$ and $n$, see \cite{InverseVIsTraffic}). Assume we are given $K$ networks $(\mathcal{V}_k,\scrA_k,\mathcal{W}_k),\, k = 1,
\ldots ,K$, (as a special case, these could be $K$ replicas of the same network $\left( {\mathcal{V}, \scrA, \mathcal{W}} \right)$) and the normalized flow data $\{ \bx^k = (x_a^k/m_a^k; \, a \in
\scrA_k); \, k = 1, \ldots ,K \}$, where $k$ is the network index and $x_a^k$ (resp., $m_a^k$) is the flow (resp., capacity) for link $a \in \scrA_k$ correspondingly.  Let $M =
\sum\nolimits_{k = 1}^K |\scrA_k|$ and $\bz = (z_1,\ldots,z_M) \defeq (
{{(\bx^{1})'}, \ldots ,{(\bx^{K})'}})$. Take the kernel matrix
to be $\boldsymbol{\Phi} = \left[ {{\phi} \left(
	{{z_i},{z_j}} \right)} \right]_{i,j = 1}^{M}$. By \cite[Thm. 2]{bertsimas2014data}, we
reformulate the inverse VI problem \eqref{cdc16-inverseVI1} as the following
Quadratic Programming (QP) problem:
\begin{align}
\mathop {\min }\limits_{\boldsymbol{\alpha}, \textbf{y}, \boldsymbol{\epsilon}} {\text{ }} &\boldsymbol{\alpha} '\boldsymbol{\Phi}\boldsymbol{\alpha}  + \gamma \left\| \boldsymbol{\epsilon}  \right\| \label{cdc16-inVI_1} \\
\text{s.t.}{\text{  }}&\textbf{e}'_{a}\textbf{N}_k '{\textbf{y}^\textbf{w}} \leq {t^{0}_a}\boldsymbol{\alpha}' \boldsymbol{\Phi}\textbf{e}_{a}^{}, \notag \\ &~~~~~~~~~~ \forall \textbf{w} \in {\mathcal{W}_k}, ~a \in {\scrA_k},~k = 1, \ldots ,K,  \notag \\
&\boldsymbol{\alpha}' \boldsymbol{\Phi}\textbf{e}_{a}^{} \leq \boldsymbol{\alpha}' \boldsymbol{\Phi}\textbf{e}_{{\tilde a}}^{}, \notag \\ &~~~~~~~~~~ \forall a,\, {\tilde a} \in { \bigcup\nolimits_{k =
		1}^K {{\scrA_k}}}
~{\text{ s}}{\text{.t}}{\text{. }}\frac{{{x}_{a}}}{{{m_{a}}}} \leq \frac{{{x}_{{\tilde a}}}}{{{m_{\tilde a}}}},  \notag \\
&\sum\limits_{a \in {\scrA_k}} {{t^{0}_a}x_{{a}}\boldsymbol{\alpha} '\boldsymbol{\Phi}\textbf{e}_{a}^{}}  - \sum\limits_{\textbf{w} \in {\mathcal{W}_k}} {{{\left( {{\textbf{d}^\textbf{w}}} \right)}' }{\textbf{y}^\textbf{w}}}  \leq {\epsilon _k}, \notag \\ &~~~~~~~~~~ \forall k = 1, \ldots ,K,  \notag \\
&\epsilon _k > 0, \quad \forall k = 1, \ldots ,K,  \notag \\
&\boldsymbol{\alpha} '\boldsymbol{\Phi}\textbf{e}_{a_0}^{} = 1, \notag 
\end{align}
where $\boldsymbol{\alpha}$, $\by = \left( {{\by^{\bw}};\bw \in {\scrW_k},k = 1, \ldots ,K} \right)$, and $\boldsymbol{\epsilon} = (\epsilon_k; k =1, \ldots, K)$ are decision vectors ($\by^{\bw}$ is the ``price'' of $\bd^{\bw}$, in particular), $\gamma$
is a regularization parameter, and $a_0$ is some arbitrary link chosen for
normalization purposes. The first constraint is for dual feasibility. The second constraint forces the
function $f(\cdot)$ (the cornerstone of the travel latency cost function $\bt(\cdot)$) to be non-decreasing on ${ \bigcup\nolimits_{k =
		1}^K {{\scrA_k}}}$. The third constraint is the primal-dual gap constraint.

Theoretically, we can derive an estimator $\bar{f}(\cdot)$ of $f(\cdot)$ by solving the QP
\eqref{cdc16-inVI_1}, thereby obtaining an optimal $\boldsymbol{\alpha}^*$. Writing ${\boldsymbol{\alpha} ^*} = (\alpha _1^{*},\ldots,
\alpha _M^{*})$, then, by \cite[Thm. 4]{bertsimas2014data}, we obtain
\begin{align}
\bar f\left( \cdot \right) = \sum\nolimits_{m = 1}^M { {\alpha
		_m^{*}\boldsymbol{\phi} ( {z_m, \cdot } )} }.  \label{cdc16-costExp}
\end{align}
However, directly solving \eqref{cdc16-inVI_1} would typically encounter numerical difficulties, due to the low rank of the kernel matrix $\bPhi$ (typically in networks with many links).
Thus, in our implementation, we will consider another formulation, by directly using the polynomial base functions of the same RKHS.  
In particular, by \cite[(3.2), (3.3), and (3.6)]{evgeniou2000regularization}, writing 
\[\phi\left( {x,y} \right) = {\left( {c + xy} \right)^n} = \sum\limits_{i = 0}^n {{n \choose i}{c^{n - i}}{x^i}{y^i}}, \]
we reformulate \eqref{cdc16-inVI_1} as
\begin{align}
\mathop {\min }\limits_{\bbeta ,\by,\boldsymbol{\epsilon} } {\text{  }}&\sum\limits_{i = 0}^n {\frac{{\beta _i^2}}{{{n \choose i}{c^{n - i}}}}}  + \gamma \left\| \boldsymbol{\epsilon}  \right\| \label{inverVI2} \\
\text{s.t.}{\text{  }}&{\textbf{e}}_a'\bN_k'{\by^\bw} \leq t_a^0{\sum\limits_{i = 0}^n {{\beta _i}\left( {\frac{{{x_a}}}{{{m_a}}}} \right)} ^i}, \; \notag \\ &~~~~~~~~~~\forall \bw \in {\mathcal{W}_k}, ~a \in {\scrA_k},~k = 1, \ldots ,K, \notag \\
&{\sum\limits_{i = 0}^n {{\beta _i}\left( {\frac{{{x_a}}}{{{m_a}}}} \right)} ^i} \leq {\sum\limits_{i = 0}^n {{\beta _i}\left( {\frac{{{x_{\tilde a}}}}{{{m_{\tilde a}}}}} \right)} ^i},\;\notag \\ &~~~~~~~~~~\forall a,~\tilde a \in {\scrA_0}~\text{s.t. }\frac{{{x_a}}}{{{m_a}}} \leq \frac{{{x_{\tilde a}}}}{{{m_{\tilde a}}}}, \notag \\
&\sum\limits_{a \in {\scrA_k}}{t_a^0{x_a}{\sum\limits_{i = 0}^n {{\beta _i}\left( {\frac{{{x_a}}}{{{m_a}}}} \right)} ^i}} - \sum\limits_{\bw \in {\mathcal{W}_k}} {\left( {{\bd^\bw}} \right)'{\by^\bw}}  \leq {\epsilon _k},\;\notag \\ &~~~~~~~~~~\forall k = 1, \ldots ,K, \notag \\
&\epsilon _k > 0, \quad \forall k = 1, \ldots ,K,  \notag \\
&{\beta _0} = 1, \notag 
\end{align}
where the last constraint is for normalization purposes, which makes sense, because an updated estimator $\hat f(\cdot)$ (as opposed to $\bar f(\cdot)$ in \eqref{cdc16-costExp}) should at least satisfy $\hat f(0) = 1$ (corresponding to the free-flow travel time; see \eqref{cdc16-costf}).
Assuming an optimal ${\bbeta ^*} = \left( {\beta _i^*;\,i = 0,1, \ldots ,n} \right)$ is obtained by solving \eqref{inverVI2}, then our final estimator for $f(\cdot)$ is
\begin{align}
\hat f\left( x \right) = \sum\limits_{i = 0}^n {\beta _i^*{x^i}}  = 1 + \sum\limits_{i = 1}^n {\beta _i^*{x^i}}.  \label{cdc16-costEstimator}
\end{align}

\subsection{OD demand estimation}

Given network $\left( {\mathcal{V}, \scrA, \mathcal{W}} \right)$, to estimate an initial OD demand matrix, we borrow the Generalized Least Squares (GLS) method proposed in \cite{hazelton2000estimation}, which assumes that the transportation network $\left( {\mathcal{V}, \scrA, \mathcal{W}} \right)$ is uncongested (in other words, for each OD pair the route choice probabilities are independent of traffic flow), and that the OD trips (traffic counts) are Poisson distributed. Note that such assumptions are a bit strong and we will relax them when finalizing our OD demand estimator by performing an adjustment procedure.

Let $\{ {{\textbf{x}^{\left( k \right)}}};
\, k = 1, \ldots, K\}$ denote $K$ observations of the flow vector
and ${\bar {\textbf{x}}}$ the average. Denote by $\bS = (1/(K - 1))\sum\nolimits_{k = 1}^K {\big(
	{{\textbf{x}^{\left( k \right)}} - \bar {\textbf{x}}} \big){{\big(
			{{\textbf{x}^{\left( k \right)}} - \bar {\textbf{x}}} \big)}' }}$
the sample covariance matrix. Let $\bP = \left[ {{p_{ir}}} \right]$ be the
route choice probability matrix, where $p_{ir}$ is the probability that a traveler between OD pair $i$
uses route $r$. Vectorize the OD demand matrix as $\bg $ and let $\bxi  = \bP'\bg$. Then, the GLS method proceeds as follows:

(\rmnum{1}) Find feasible routes for each OD pair, thus obtaining the link-route incidence matrix $\bA$. 

(\rmnum{2}) 
Solve sequentially the following two problems:
\begin{align}
\text{(P1)} \quad \mathop {\min }\limits_{\bxi  \succeq \textbf{0}} \quad \frac{K}{2}\bxi '\bQ\bxi  - \bb'\bxi,   \label{qp2}
\end{align}
where $\bQ = \bA'{\bS^{ - 1}}\bA$ and $\textbf{b} =
\sum\nolimits_{k = 1}^K {\bA'{\bS^{ - 1}}{\textbf{x}^{\left( k
			\right)}}} $, and
\begin{align}
\text{(P2)} \quad &\mathop {\min }\limits_{\bP \succeq \textbf{0}, \,\bg \succeq \textbf{0} } \quad h\left( {\bP,\bg } \right) \label{qp3} \\
\text{s.t.} \quad 
& p_{ir} = 0 \quad \forall (i,r) \in \{(i,r): r \notin \scrR_i\}, \notag \\
&\bP'\bg  = {\bxi ^*}, \notag \\
&\bP\textbf{1} = \textbf{1}, \notag 
\end{align}
where $h(\bP, \bg)$ can be taken as any smooth scalar-valued function, $\bxi^*$ is the optimal solution to (P1), and $\textbf{1}$ denotes the vector with all $1$'s as its entries. Note that (P1) (resp., (P2)) is a typical \textit{Quadratic Programing (QP)} (resp., \textit{Quadratically Constrained Programming (QCP)}) problem. Letting $(\bP^*, \bg^*)$ be an optimal solution to (P2), then $\bg^*$ is our initial estimate of the demand vector.

\begin{rmk} \label{rem:od}
	\em{We note here that the GLS method would encounter numerical difficulties when the network size is large, because there would be too many decision variables in (P2). Note also that the GLS method is valid under an ``uncongestion'' assumption and, to take the congestion, here the link flows, into account, we in turn consider a bi-level optimization problem in the following.}
\end{rmk}

Assume now the function $f(\cdot)$ in \eqref{cdc16-costf} is available. For any given feasible $\bg$ ($\succeq \textbf{0}$), let $\bx(\bg)$ be the optimal solution to the TAP \eqref{cdc16-tap}.
In the following, assume $\tilde{\bx} = (\tilde{x}_a; \;a \in \scrA)$ to be the observed flow vector. Assuming we are given an initial demand vector $\bg^{0}$, we consider the following bi-level optimization problem (BiLev) (see \cite{spiess1990gradient} \& \cite{lundgren2008heuristic}):

\begin{align}
\text{(BiLev)} \quad \mathop {\min }\limits_{\bg \succeq \textbf{0}} {\text{  }}F\left( \bg \right) \defeq \sum\limits_{a \in \scrA} {{{\left( {{x_a}\left( \bg \right) - {{\tilde x}_a}} \right)}^2}}. \label{bilevel} \end{align}

Note that $F\left( \bg \right)$ has a lower bound $0$ which guarantees the convergence of the algorithm (see Alg. \ref{alg:demandAdjustment}) that we are going to apply.


To solve the (BiLev) numerically, thus adjusting the demand vector iteratively, we leverage a gradient-based algorithm (Alg. \ref{alg:demandAdjustment}). In particular, suppose that the route probabilities are locally constant. For OD pair $i$, considering only the shortest route $r_i(\bg)$, we have (see \cite{spiess1990gradient})
\begin{align}
\frac{{\partial {x_a}\left( \bg  \right)}}{{\partial {g _i}}} = {\delta _{{r_i(\bg)}a}} = \left\{ \begin{gathered}
1,{\text{ if }}a \in {r_i(\bg)}, \hfill \\
0,{\text{ otherwise}}{\text{.}} \hfill \\  
\end{gathered}  \right. \label{jacobbb}
\end{align}
Thus, by \eqref{jacobbb} we obtain the Jacobian matrix
\begin{align}
\left[ {\frac{{\partial {x_a}\left( {{\bg}} \right)}}{{\partial {g _i}}};\;a \in \scrA,\;i \in \left\{ {1, \ldots ,\left| \scrW \right|} \right\}} \right]. \label{jacoMat}
\end{align}

\begin{rmk} \label{rem:shortest}
	\em{There are three reasons why we consider only the shortest routes for the purpose of calculating the Jacobian:
		(\rmnum{1}) GPS navigation is widely-used by vehicle drivers so that they tend to always select the shortest routes between their OD pairs. (\rmnum{2}) There are many efficient algorithms for finding the shortest route for each OD pair. (\rmnum{3}) If considering more than one route for an OD pair, then the route flows cannot be uniquely determined by solving the TAP \eqref{cdc16-tap}, thus leading to unstable route-choice probabilities, which would undermine the accuracy of the approximation to the Jacobian matrix in \eqref{jacoMat}.}
\end{rmk}

Let us now compute the gradient of $F\left( \bg \right)$. We have 
\begin{align}
&\nabla F\left( \bg  \right) = \left( {\frac{{\partial F\left( \bg  \right)}}{{\partial {g _i}}}};\;i = 1, \ldots ,\left| \scrW \right| \right)  \notag \\
&= \left( {\sum\limits_{a \in \scrA} {2\left( {{x_a}\left( \bg  \right) - {{\tilde x}_a}} \right)\frac{{\partial {x_a}\left( \bg  \right)}}{{\partial {g _i}}}} ;\;i = 1, \ldots ,\left| \scrW \right|} \right). \label{gradi}
\end{align}

We summarize the procedures for adjusting the OD demand matrices as Alg. \ref{alg:demandAdjustment}, whose convergence will be proven in the following proposition.
\begin{algorithm}
	\caption{Adjusting OD demand matrices}
	\label{alg:demandAdjustment}
	\begin{algorithmic}[1]
		\Require  the road network $\left( {\mathcal{V}, \scrA, \mathcal{W}} \right)$; the function $f(\cdot)$ in \eqref{cdc16-costf}; the observed flow vector $\tilde{\bx} = (\tilde{x}_a; \;a \in \scrA)$; the initial demand vector ${\bg^{0}} = (g^{0}_i; \;i = 1,\ldots,\left| \scrW \right|)$; two positive integer parameters $\rho$, $T$; two real parameters
		$\varepsilon_1 \geq 0$, $\varepsilon_2 > 0$. 
		\State \textbf{Step 0:} Initialization. Take the demand vector $\bg^{0}$ as the input, solve the TAP \eqref{cdc16-tap} using Alg. \ref{alg:msa} to obtain $\bx^0$. Set $l=0$. If $F\left(\bg^{0}\right) = 0$, stop; otherwise, go onto Step 1.
		\State \textbf{Step 1:} Computation of a descent direction. Calculate ${\bh^l} =  - \nabla F\left( {{\bg ^l}} \right)$ by \eqref{gradi}.
		\State \textbf{Step 2:} Calculation of a search direction. For $i = 1, \ldots ,\left| \scrW \right|$ set 
		$$\bar h_i^l = \left\{ \begin{gathered}
		h_i^l,{\text{  if }}\left( {g _i^l > \varepsilon_1 } \right){\text{ or }}\left( {g _i^l \leq \varepsilon_1 {\text{ and }}h_i^l > 0} \right){\text{, }} \hfill \\
		0,{\text{  otherwise.}} \hfill \\ 
		\end{gathered}  \right.$$
		
		\State \textbf{Step 3:} Armijo-type line search: 
		\begin{itemize}
			\item 		\textbf{3.1:} Calculate the maximum possible step-size $\theta _{\max }^l = \min \left\{ { - \frac{{g _i^l}}{{\bar h_i^l}}; \; \bar h_i^l < 0, i = 1, \ldots ,\left| \scrW \right|} \right\}$. 
			\item 		\textbf{3.2:} Determine ${\theta ^l} = \mathop {\arg \min }\limits_{\theta  \in \scrS} F\left( {{\bg ^l} + \theta {{\bar {\bh}}^l}} \right)$, where $\scrS \defeq \left\{{\theta _{\max }^l,{{\theta _{\max }^l} \mathord{\left/
						{\vphantom {{\theta _{\max }^l} \rho }} \right.
						\kern-\nulldelimiterspace} \rho },{{\theta _{\max }^l} \mathord{\left/
						{\vphantom {{\theta _{\max }^l} {{\rho ^2}}}} \right.
						\kern-\nulldelimiterspace} {{\rho ^2}}}, \ldots ,{{\theta _{\max }^l} \mathord{\left/
						{\vphantom {{\theta _{\max }^l} {{\rho ^{T - 1}}}}} \right.
						\kern-\nulldelimiterspace} {{\rho ^{T}}}}}, 0 \right\}$.
		\end{itemize}

		\State \textbf{Step 4:} Update and termination. 
		\begin{itemize}
			\item 		\textbf{4.1:} Set ${\bg ^{l + 1}} = {\bg ^l} + {\theta ^l}{{\bar \bh}^l}$. Using $\bg^{{l+1}}$ as the input, solve the TAP \eqref{cdc16-tap} to obtain $\bx^{{l+1}}$.
			\item 		\textbf{4.2:} If $\frac{{F\left( {{\bg^l}} \right) - F\left( {{\bg^{l + 1}}} \right)}}{{F\left( {{\bg^0}} \right)}} < {\varepsilon _2}$, stop the iteration; otherwise, go onto Step 4.3. 
			\item 		\textbf{4.3:} Set $l=l+1$ and return to Step 1.	
		\end{itemize}
		
	\end{algorithmic}
\end{algorithm}

\begin{algorithm}
	\caption{Method of Successive Averages (MSA) (see \cite{noriega2007algorithmic})}
	\label{alg:msa}
	\begin{algorithmic}[1]
		\Require  the road network $\left( {\mathcal{V}, \scrA, \mathcal{W}} \right)$; the function $f(\cdot)$ in \eqref{cdc16-costf}; the demand vector ${\bg} = (g_i; \;i = 1,\ldots,\left| \scrW \right|)$; a real parameter $\varepsilon > 0$. 
		\State \textbf{Step 0:} Initialization. Initialize link flows: $x_a^{\ell} = 0$ for $a \in \scrA$; set iteration counter $\ell=0$.
		\State \textbf{Step 1:} Compute new extremal flow. Set $\ell = \ell + 1$.
		
		\begin{itemize}
			\item \textbf{1.1:} Update link travel times (costs) based on current link flows: $t_a^{\ell} = t_a\left( {x_a^{\ell - 1}} \right)$.
			
			\item \textbf{1.2:} Carry out ``all-or-nothing'' assignment of the demands $\bg$ on current shortest paths to obtain $y_a^{\ell}$.
			
		\end{itemize}
		\State \textbf{Step 2:} Update link flows.
		
		Update link flows by using the MSA step size:
		\[x_a^{\ell} = x_a^{\ell - 1} + {\lambda ^\ell}\left( {y_a^{\ell} - x_a^{\ell - 1}} \right),\]
		where ${\lambda ^\ell} = {1 \mathord{\left/
				{\vphantom {1 l}} \right.
				\kern-\nulldelimiterspace} \ell}$.
		
		\State \textbf{Step 3:} Stopping criterion (slightly different than that in \cite{noriega2007algorithmic}).
		
		Compute the \textit{Relative Gap}, $RG$, as:
		\[RG = \frac{{{{\left\| {{\bx^{\ell}} - {\bx^{\ell-1}}} \right\|}}}}{{{{\left\| {{\bx^{\ell}}} \right\|}}}}.\]

		If $RG < \varepsilon$ terminate; otherwise return to Step 1.
	\end{algorithmic}
\end{algorithm}


\begin{prop} \label{prop1} Alg. \ref{alg:demandAdjustment} converges.
\end{prop}

\begin{pf}
	If the initial demand vector $\bg^{0}$ satisfies $F\left(\bg^{0}\right) = 0$, then, by Step 0, the algorithm stops (trivial case). Otherwise, we have $F\left(\bg^{0}\right) > 0$, and it is seen from \eqref{bilevel} that the objective function $F\left(\bg\right)$ has a lower bound $0$. In addition, by the line search and the update steps (Steps 3.2 and 4.1, in particular), we obtain 
		\begin{align}
		F\left( {{\bg^{l + 1}}} \right) = F\left( {{\bg^l} + {\theta ^l}{{\bar \bh}^l}} \right) = \mathop {\min }\limits_{\theta  \in \scrS} F\left( {{\bg^l} + \theta {{\bar \bh}^l}} \right) \leq F\left( {{\bg^l}} \right), \forall l, \notag
		\end{align}
		where the last inequality holds due to $0 \in \scrS$, indicating that the nonnegative objective function in \eqref{bilevel} is non-increasing as the number of iterations increases. Thus, by the well-known monotone convergence theorem, the convergence of the algorithm can be guaranteed. 
		\hfill $\qed$
\end{pf}

\begin{rmk}
	\em{Alg. \ref{alg:demandAdjustment} is a variant of the algorithms proposed in \cite{spiess1990gradient} \& \cite{lundgren2008heuristic}. We use a different method to calculate the \textit{step-sizes} (resp., \textit{Jacobian matrix}) than that in \cite{spiess1990gradient} (resp., \cite{lundgren2008heuristic}). Moreover, as a subroutine, Alg. \ref{alg:msa} is borrowed from \cite{noriega2007algorithmic}, which has the advantage of being easy to implement.}
\end{rmk}

\subsection{Price of anarchy} \label{sec:poa}

As discussed in Sec. \ref{sec:intro}, one of our goals is to measure inefficiency in the traffic network performance due to the non-cooperative behavior of drivers. Thus, we compare the network performance under a user-centric policy vs. a system-centric policy. As a metric for this comparison  we define the POA as the ratio
between the total latency, i.e., the total travel time over all drivers
in different OD pairs, obtained under Wardrop flows (user-centric policy) and that obtained
under socially optimal flows (system-centric policy). 

Given network $\left( {\mathcal{V}, \scrA, \mathcal{W}} \right)$,
as in \cite{CDC16}, we define its \textit{total latency} as
\begin{align}
L(\bx) =\sum_{a \in \scrA}x_a t_a(x_a). \label{TotalLatency2}
\end{align}
Then, the socially optimum flow vector, denoted by $\bx^*$, is the solution to the following
Non-Linear Programming (NLP) problem (see \cite{patriksson1994traffic} \& \cite{Sepid2016}):
\begin{align}
\mathop {\min }\limits_{{\bx \in \scrF}} {\text{ }} 
\sum_{a \in \scrA}x_a t_a(x_a).  \label{obj_soc}
\end{align}
Note that here the objective function is different from the one we use in
the TAP (see Sec. \ref{sec:forward}); for a
detailed explanation, see \cite{dafermos1969traffic}.
Recalling that we consider the total latency as the performance metric, we define the \textit{Price of Anarchy} as
\begin{equation}
\text{POA} \defeq \dfrac{L(\bx^{ne})}{L(\bx^*)} = \frac{\sum_{a \in
		\scrA}x_a^{ne}t_a(x_a^{ne})}{\sum_{a \in
		\scrA}x_a^*t_a(x_a^*)},
\end{equation}
where $x_a^{ne}$ denotes the Wardrop's equilibrium flow on link $a \in \mathcal{A}$ estimated from the speed dataset (for details refer to \cite{CDC16}), and $x_a^{*}$ is the socially optimal flow on link $a \in \mathcal{A}$ obtained by solving (\ref{obj_soc}), which requires the travel latency cost functions $t_a(\cdot)$ that are estimated by solving the inverse problem (see Sec. \ref{cdc16-sec InverseVI-uni}). Thus, POA quantifies the inefficiency that a society has to deal with due to non-cooperative behavior of its members.

\subsection{Sensitivity analysis}

Write ${\bt^{0}} \defeq \left(
{{t^{0}_a; \, a \in \scrA}} \right)$, $\bm
\defeq \left( {{m_a; \, a \in \scrA}} \right)$, and
\begin{align}
V\left( {{\bt^{0}},{\bm}} \right) \defeq \mathop {\min }\limits_{{\bx \in \scrF}} \sum\limits_{a \in \scrA} {\int_0^{{x_a}} {{t^{0}_a}f\left( {\frac{s}{{{m_a}}}} \right)ds} }.
\end{align}

Let ${\bx^*} \defeq \left( {x_a^*; \, a \in \scrA} \right)$
denote the solution to \eqref{cdc16-tap}. Using
\cite[Thm. 19.5]{simon1994mathematics}, we obtain
\begin{align}
\frac{{\partial V\left( {{\bt^{0}},\bm} \right)}}{{\partial {t^{0}_a}}} &= \int_0^{x_a^*} {f\left( {\frac{s}{{{m_a}}}} \right)ds} , \label{cdc16-s1} \\
\frac{{\partial V\left( {{\bt^{0}},\bm} \right)}}{{\partial {m_a}}} &= \int_0^{x_a^*} {{t^{0}_a}\dot f\left( {\frac{s}{{{m_a}}}} \right)\left( { - \frac{s}{{m_a^2}}} \right)ds}, \label{cdc16-s2}
\end{align}
where $\dot f(\cdot)$ denotes the derivative of $f(\cdot)$.

\section{Dataset Description and Processing} \label{sec: dataset}

\subsection{Description of the Eastern Massachusetts dataset} \label{dataEMA}

\begin{figure}[htb]
	\centering
	\includegraphics[height=5cm]{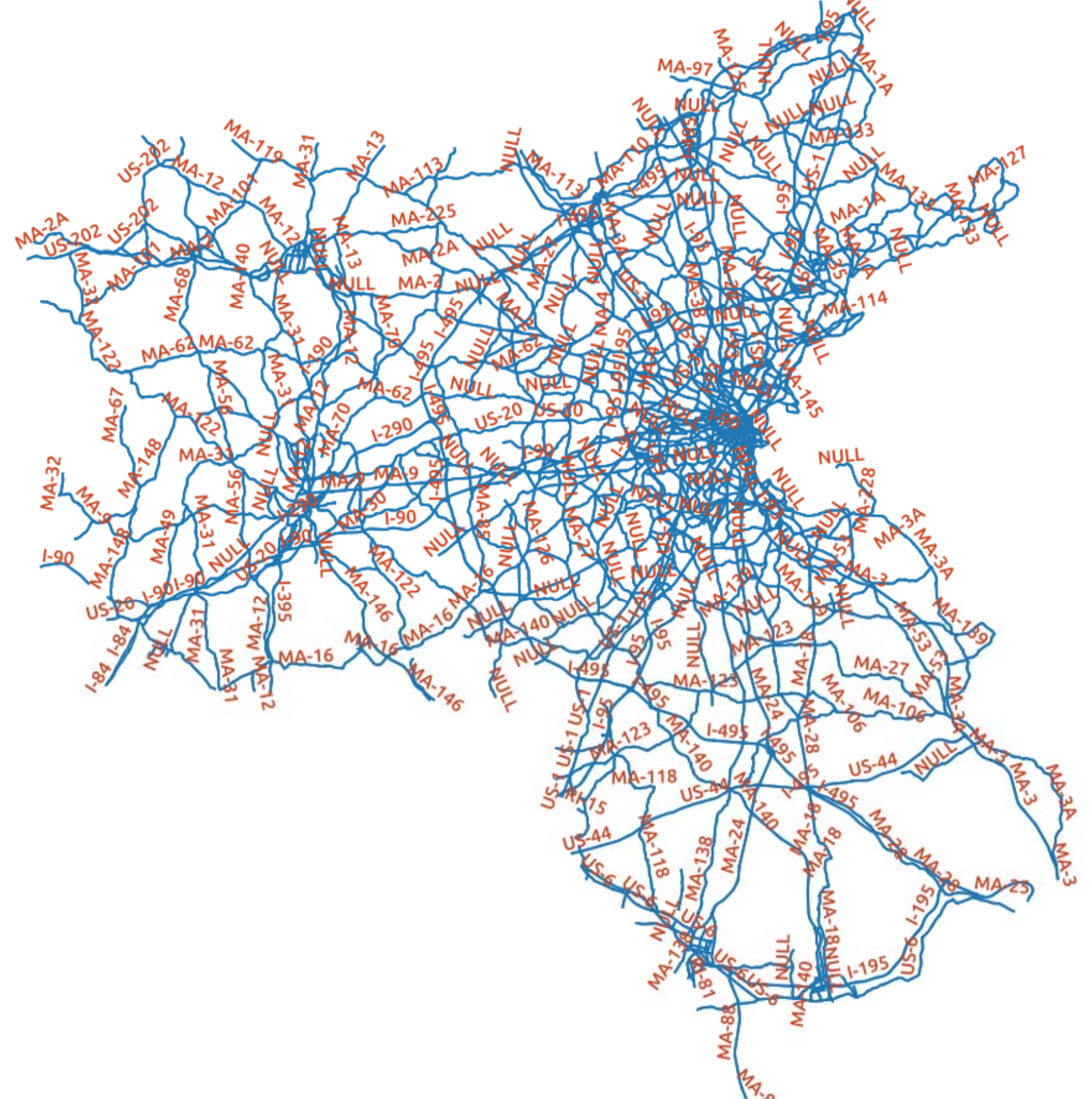}
	\caption{All available road segments in Eastern Massachusetts (from \cite{CDC16}).}
	\label{eastMA}
\end{figure}

We deal with two datasets for the Eastern Massachusetts (EMA) road network: 
\noindent (\rmnum{1}) The speed dataset, provided to us by the Boston Region Metropolitan Planning Organization (MPO), includes the spatial average speeds for more than 13,000 
road segments (with the average distance of 0.7 miles; see Fig. \ref{eastMA}) of EMA, covering the average speed for every minute of the year 2012.
For each road segment, identified with a unique {\em tmc (traffic message
	channel)} code, the dataset provides information such as speed data
(instantaneous, average and free-flow speed) in mph, date and time, and
traveling time (minute) through that segment. 
\noindent (\rmnum{2})
The flow capacity (\# of vehicles per hour) dataset, also provided by the MPO, includes
capacity data -- vehicle counts for each road segment -- for more than 100,000
road segments (average distance of 0.13 miles) in EMA. For more detailed information of these two datasets, see \cite{CDC16}.

\begin{figure}[htb]
	\centering
	\subfigure[(from \cite{CDC16})]{\label{cdc16-sub-net-a}
		\includegraphics[height=4.25cm]{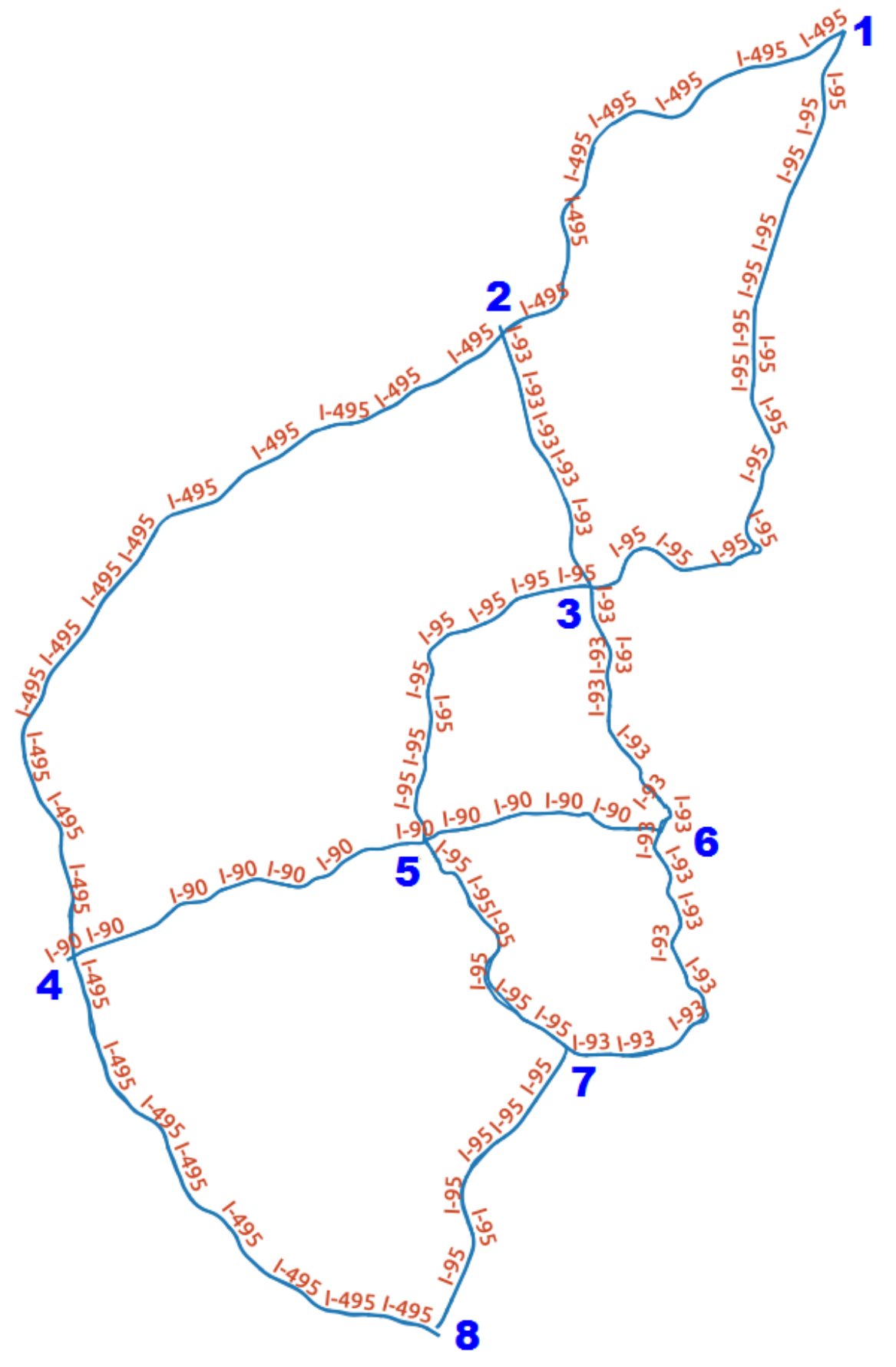} }
	\subfigure[(from \cite{CDC16})]{\label{cdc16-sub-net-b}
		\includegraphics[height=4.25cm]{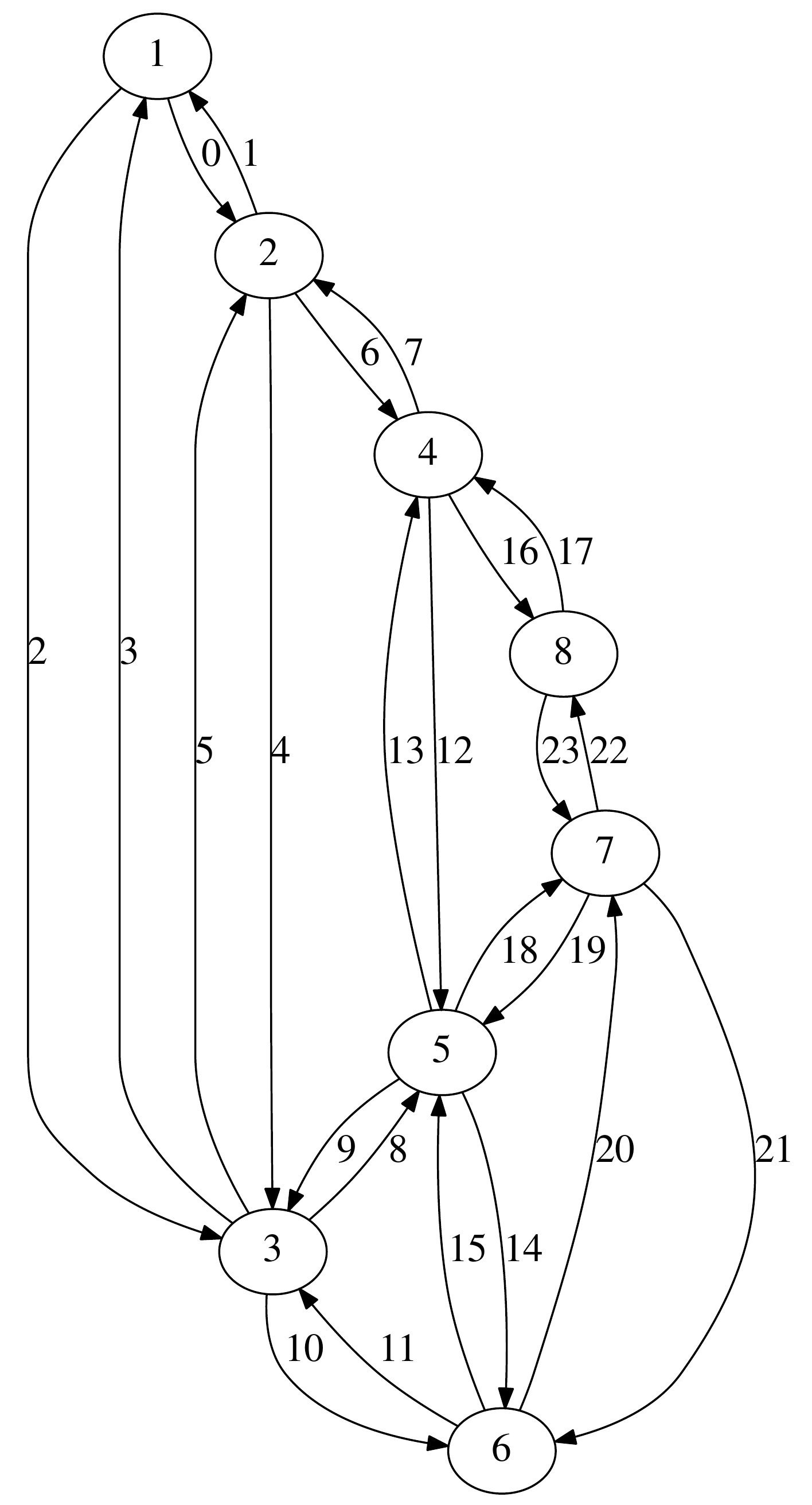} }
	\subfigure[]{\label{sub-net-a-ext}
		\includegraphics[height=4.25cm]{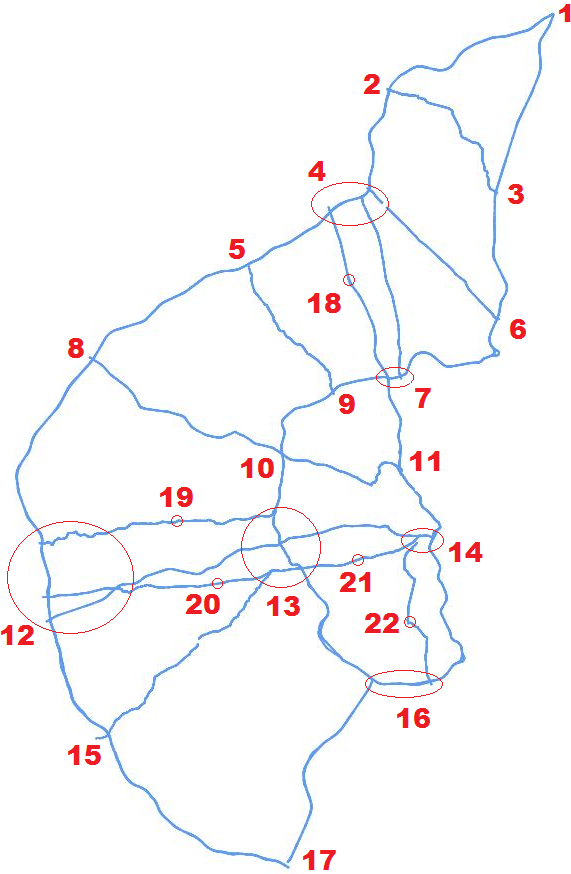} }
	\subfigure[]{\label{sub-net-b-ext}
		\includegraphics[height=5cm]{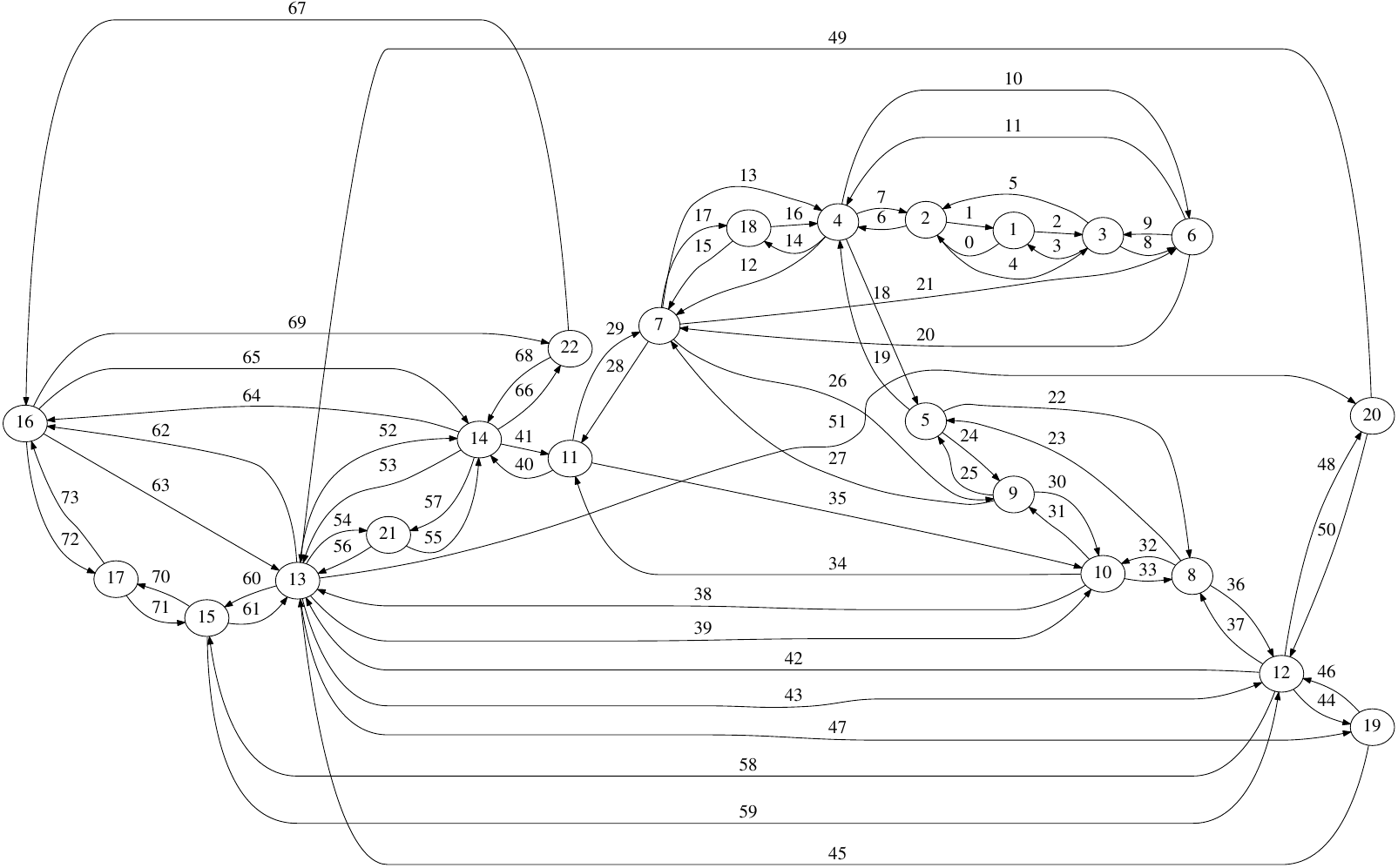} }
	\caption{(a) An interstate highway sub-network of EMA (the blue numbers indicate node indices); (b)
		The topology of the sub-network in Fig. \ref{cdc16-sub-net-a} (the numbers beside arrows are link indices,
		and the numbers inside ellipses are node indices); (c) An extended highway sub-network of EMA (the red numbers indicate node indices); (d)
		The topology of the sub-network in Fig. \ref{sub-net-a-ext}.}
	\label{cdc16-sub-net}
\end{figure}

\subsection{Preprocessing} \label{prep}

Other than selecting an extended subnetwork (shown in Fig. \ref{sub-net-a-ext} as opposed to the smaller subnetwork Fig. \ref{cdc16-sub-net-a}) of the EMA road network (shown in Fig. \ref{eastMA}), we perform exactly the same preprocessing procedures as those in \cite{CDC16} on the datasets mentioned in Sec. \ref{dataEMA}. Thus, we end up with traffic flow data and road (link) parameters (\textit{flow capacity} and \textit{free-flow travel time}) for the two subnetworks (shown in Figs. \ref{cdc16-sub-net-a} and \ref{sub-net-a-ext} respectively; from now on, let us denote them by $\scrI_1$ and $\scrI_2$ respectively), where $\scrI_1$ contains only the interstate highways and $\scrI_2$ also contains the state highways in EMA. Note that $\scrI_1$ (resp., $\scrI_2$) consists of 8 (resp., 22) nodes and 24 (resp., 74) links; their topologies are shown in Figs. \ref{cdc16-sub-net-b} and \ref{sub-net-b-ext} respectively. Assuming that each node could be an origin and a destination, then there are $8 \times (8-1) = 56$ (resp., $22 \times (22-1) = 462$) OD pairs in $\scrI_1$ (resp., $\scrI_2$).

\subsection{Estimating initial OD demand matrices} \label{sec:ODmat}

Operating on $\scrI_1$, we solve the QP (P1) and the QCP (P2) using data corresponding to five different
time periods (AM, MD, PM, NT, and weekend) of four months (Jan., Apr.,
Jul., and Oct.) in 2012, thus obtaining 20 different OD demand matrices
for these scenarios. 
As noted in \cite{CDC16}, the GLS method assumes the traffic network to be uncongested, leading to the fact that the estimated OD demand matrices for non-peak periods (MD/NT/weekend) are relatively more accurate than those for peak periods (AM/PM). 
After obtaining estimates for travel latency cost functions in Sec. \ref{cost}, based on the estimates for the OD demand matrices of $\scrI_1$, we will conduct a demand adjustment procedure for $\scrI_2$ in Sec. \ref{od-adj}.

\subsection{Estimating cost functions} \label{cost}

Operating on $\scrI_1$ using the flow data and the OD demand matrices obtained in Secs. \ref{prep} and \ref{sec:ODmat} respectively, we estimate the travel latency cost functions for 20 different scenarios, via the estimator \eqref{cdc16-costEstimator}, by solving the QP
\eqref{inverVI2} accordingly. As in \cite{InverseVIsTraffic}, to make the estimates reliable, for each
scenario, we perform a 3-fold cross-validation.
Note that \cite{InverseVIsTraffic} applied the alternative estimator \eqref{cdc16-costExp}, which is numerically not as stable.

\subsection{Sensitivity analysis}

To illustrate our method of analyzing sensitivities for the TAP, we again conduct the numerical experiments on $\scrI_1$. Take the scenario corresponding to the PM peak period on
January 10, 2012. The results will be shown in Sec. \ref{sec:sensResult}.

\subsection{OD demand adjustments} \label{od-adj}

We first demonstrate the effectiveness of the demand-adjusting algorithm (Alg. \ref{alg:demandAdjustment}) using the Sioux-Falls benchmark dataset (see \cite{BarGera16}). Then, assuming the per-road travel latency cost functions are available (we take the travel latency cost functions derived from $\scrI_1$ as in Sec. \ref{cost}), we apply Alg. \ref{alg:demandAdjustment} to $\scrI_2$, which contains $\scrI_1$ as one of its representative subnetworks. Note that the main difference between $\scrI_1$ and $\scrI_2$ is the modeling emphasis ($\scrI_1$ only takes account of the interstate highways, while $\scrI_2$ also encompasses the state highways, thus containing more details of the real road network of EMA), and we can think of $\scrI_1$ as a ``landmark'' subnetwork of $\scrI_2$. Based on the initially estimated demand matrices for $\scrI_1$, we will implement the following generic demand-adjusting scheme so as to derive the OD demand matrices for $\scrI_2$:

Given a network ($\scrI_2$ in our case) with any size, we can select its ``landmark'' subnetworks ($\scrI_1$ in our case) (based on the information of road types, pre-identified centroids, etc.) with acceptably smaller sizes, say, we end up with $N$ ($N=1$ in our case) such subnetworks. Then, for each subnetwork, we estimate its demand matrix by solving sequentially the QP (P1) and the QCP (P2) (cf. Sec. \ref{sec:ODmat}). Setting the demand for any OD pair not belonging to this subnetwork to zero, we obtain a ``rough'' initial demand matrix for the entire network ($\scrI_2$ in our case). Next, take the average of these initial demand matrices. Finally, we adjust the average demand matrix based on the flow observations of the entire network.

As noted in Remark \ref{rem:od}, the reason why we do not directly solve (P1) and (P2) for the larger network ($\scrI_2$ in our case) is that there are too many decision variables in (P2) and this would lead to numerical difficulties.

\subsection{POA evaluations}

We calculate the POA values for $\scrI_2$ for the PM period of April 2012.

\section{Numerical Results} \label{Sec:Rsults}


The results for the initial estimation of OD demand matrices and the estimation of travel latency cost functions are similar to those shown in \cite{CDC16}, and we omit them here. We will focus on presenting the results for sensitivity analysis (derived for $\scrI_1$), demand adjustment (derived for the Sioux-Falls benchmark network; note that we will not show the detailed demand adjustment results for $\scrI_2$, because we do not have the ground truth for a comparison), and the POA (derived for $\scrI_2$).

\subsection{Results from OD demand adjustment}
\begin{figure}[htb]  
	\centering
	\subfigure[Objective function value of the (BiLev) vs. \# of iterations.]{\label{objFun_Sioux}
		\includegraphics[width=0.4\textwidth]{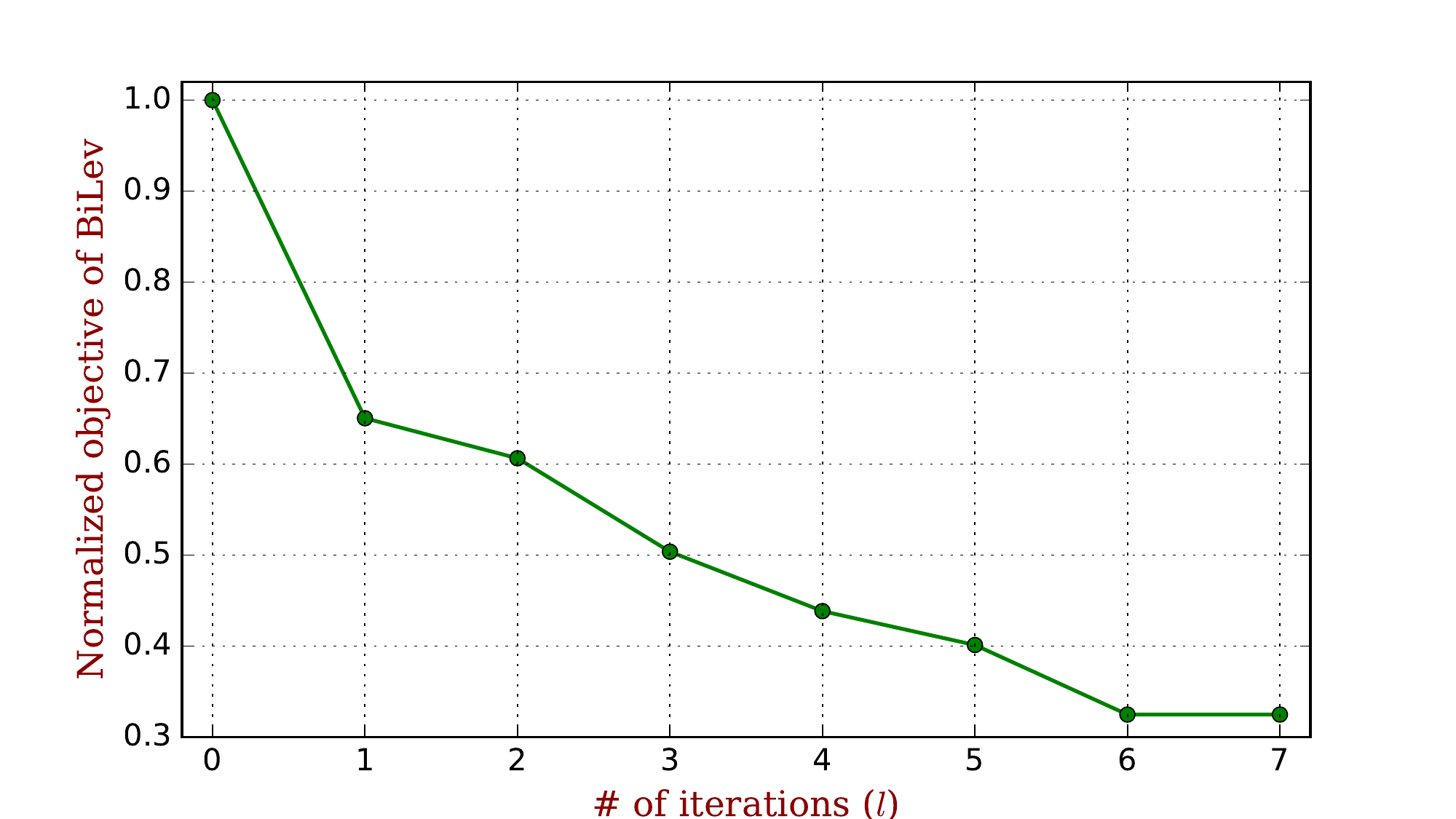}}
	\subfigure[Demand difference w.r.t. ground truth vs. \# of iterations.]{\label{demandsDiff_biLev_Sioux}
		\includegraphics[width=0.4\textwidth]{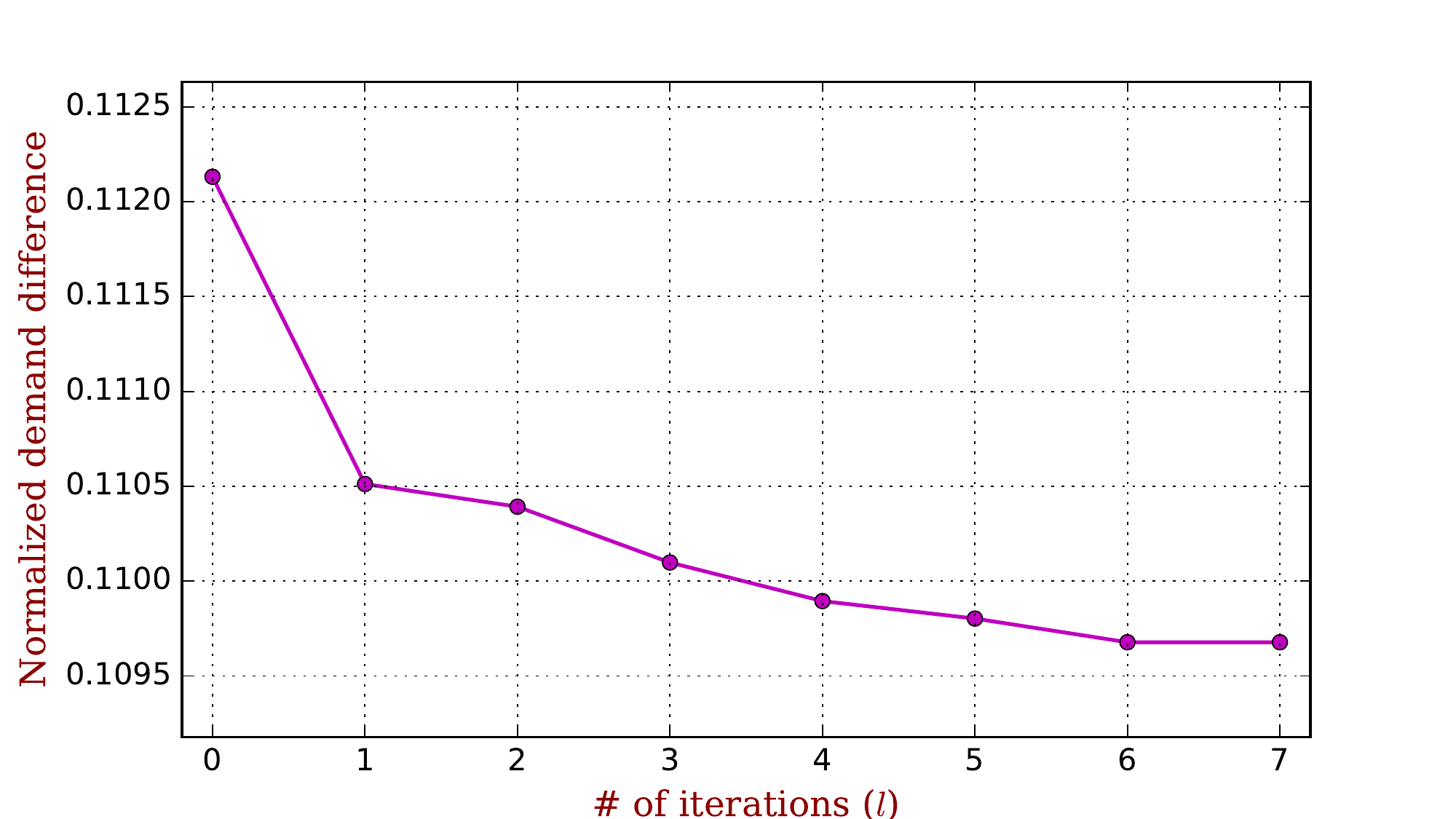}}
	\caption{Key quantities vs. \# of iterations (Sioux-Falls).}
	\label{fig:Sioux}
\end{figure}

We conduct numerical experiments on
the Sioux-Falls network, which contains 24 zones, 24 nodes, and 76 links. For each OD pair, the initial demand is taken by scaling the ground truth demand with a uniform distribution $[0.8, 1.2]$. The ground truth $f\left(\cdot\right)$ is taken as $f\left(x\right) = 1 + 0.15x^4, \; \forall x \geq 0$, and is assumed directly available.  When implementing Alg. \ref{alg:demandAdjustment} (use Alg. \ref{alg:msa} as a subroutine), we set $\rho = 2$, $T = 10$, $\varepsilon_1 =0$, $\varepsilon_2 = 10^{-20}$ and  $\varepsilon = 10^{-6}$. Fig. \ref{objFun_Sioux} shows that, after 7 iterations, the objective function value of the (BiLev) \eqref{bilevel} has been reduced by more than $65\%$. Fig. \ref{demandsDiff_biLev_Sioux} shows that, the distance between the adjusted demand and the ground truth demand keeps decreasing with the number of iterations, and the distance changes very slightly, meaning the adjustment procedure does not alter the initial demand much. Note that in Fig. \ref{objFun_Sioux}, the vertical axis corresponds to the normalized objective function value of the (BiLev), i.e., ${{F\left( {{\bg^l}} \right)} \mathord{\left/
		{\vphantom {{F\left( {{\bg^l}} \right)} {F\left( {{\bg^0}} \right)}}} \right.
		\kern-\nulldelimiterspace} {F\left( {{\bg^0}} \right)}}$ and, in Fig. \ref{demandsDiff_biLev_Sioux}, the vertical axis denotes the normalized distance between the adjusted demand vector and the ground truth, i.e., ${{\left\| {{\bg^l} - {\bg^*}} \right\|} \mathord{\left/
		{\vphantom {{\left\| {{\bg^l} - {\bg^*}} \right\|} {\left\| {{\bg^*}} \right\|}}} \right.
		\kern-\nulldelimiterspace} {\left\| {{\bg^*}} \right\|}}$, where $\bg^*$ is the ground-truth demand vector.

\subsection{Results for POA estimation}

\begin{table*}[t]
	\centering
	\caption{Scaled sensitivity analysis results.} \label{cdc16-tab1}
	\resizebox{12cm}{!}
	{
		\begin{tabular}{|l|l|l|l||l|l|l|l|}
			\hline
			\multicolumn{4}{|c||}{\(\frac{{\partial V\left( {{\bt^{{0}}},\bm} \right)}}{{\partial {t^{{0}}_a}}}\)}
			& \multicolumn{4}{|c|}{\(\frac{{\partial V\left( {{\bt^{{0}}},\bm} \right)}}{{\partial {m_a}}}\)} \\
			\hline
			(0, 0.226)  & (1, 0.177)  & (2, 0.26)  & (3, 0.208) &  (0, -0.009)  & (1, -0.005)  & (2, -0.011)  & (3, -0.005) \\
			\hline
			(4, 0.54)  & (5, 0.198) & (6, 0.199)  & (7, 0.242) & (4, -0.036)  & (5, -0.002) & (6, -0.009)  & (7, -0.017)  \\
			\hline
			(\textcolor{red}{8}, 0.924)  & (9, 0.327)  & (\textcolor{red}{10}, 0.951) & (11, 0.543) & (8, -0.496)  & (9, -0.009)  & (\textcolor{blue}{10}, -0.605) & (11, -0.094) \\
			\hline
			(12, 0.389) & (13, 0.245) & (14, 0.539) & (15, 0.42) & (12, -0.048) & (13, -0.007) & (14, -0.119) & (15, -0.103) \\
			\hline
			(16, 0.178) & (17, 0.129) & (18, 0.341) & (\textcolor{red}{19}, 1.0) & (16, -0.004) & (17, -0.002) & (18, -0.011) & (\textcolor{blue}{19}, -0.883) \\
			\hline
			(\textcolor{red}{20}, 0.892) & (\textcolor{red}{21}, 0.838) & (22, 0.234) & (23, 0.833) & (\textcolor{blue}{20}, -0.888) & (\textcolor{blue}{21}, -0.776) & (22, -0.006) & (\textcolor{blue}{23}, -1.0) \\
			\hline
	\end{tabular}}
\end{table*}

After implementing the demand adjusting scheme, we obtain the demand matrices for $\scrI_2$ on a daily-basis, as opposed to those for $\scrI_1$ on a monthly-basis. Note that, even for the same period of a day and within the same month, slight demand variations among different days are possible; thus, our POA results for $\scrI_2$ would be more accurate than those for $\scrI_1$ that have been shown in \cite{CDC16}.

\begin{figure}[hbt]
	\centering
	\includegraphics[width=0.4\textwidth]{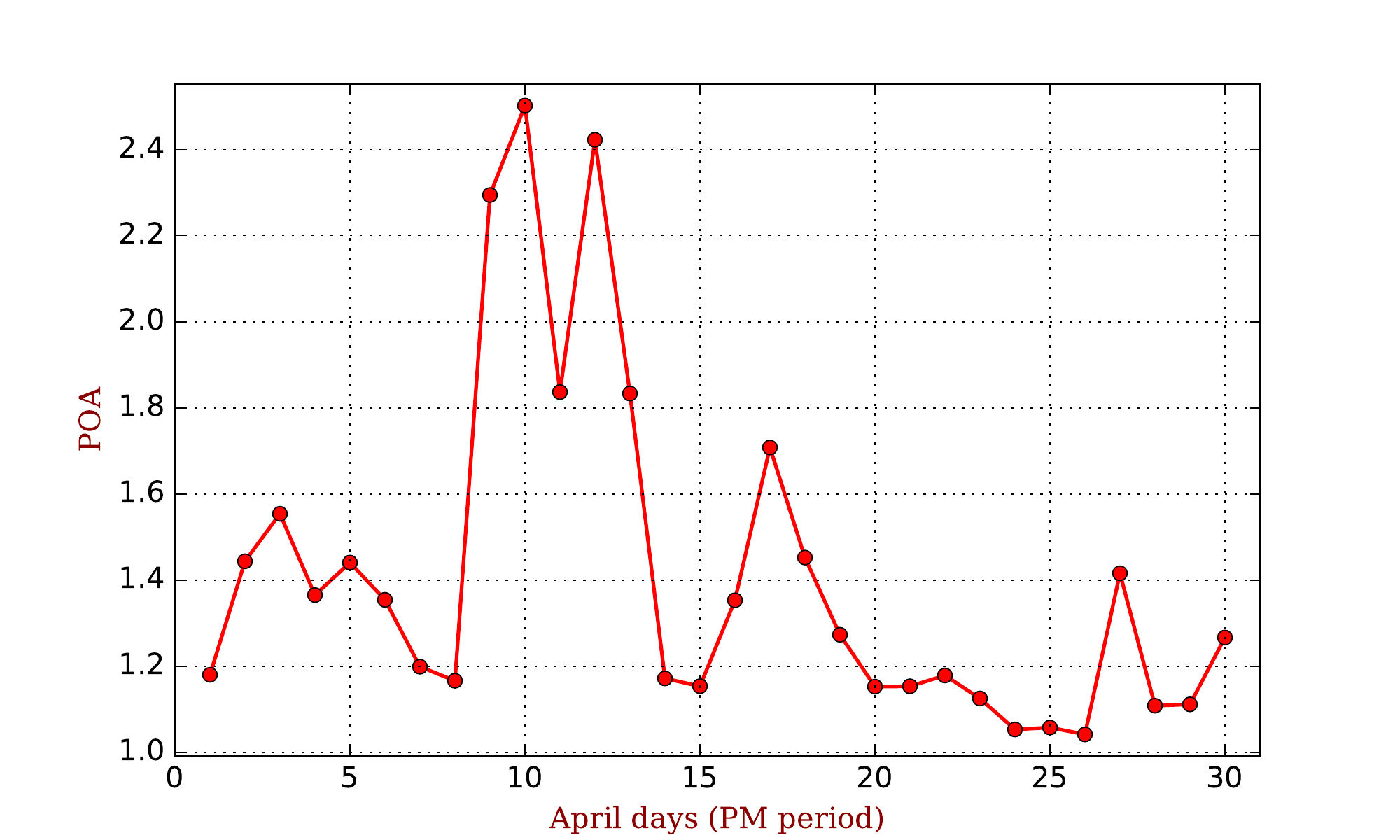}
	\caption{POA of $\scrI_2$ for PM period  in Apr. 2012 based on avg. flow on each link.}\label{IFAC-POA_PM_Apr}
\end{figure}

The POA values shown in Fig. \ref{IFAC-POA_PM_Apr} have larger variations than those for $\scrI_1$ in \cite{CDC16}; some are now closer to 1 but some go beyond 2.2, meaning we have larger potential to improve the road network. However, when taking the average of the POA values for all the 30 days, both $\scrI_2$ and $\scrI_1$ result in an average POA approximately equal to 1.5, meaning we can gain an efficiency improvement of about 50\%; thus, the results are consistent.

\subsection{Results from sensitivity analysis}  \label{sec:sensResult}

The sensitivity analysis results derived for $\scrI_1$ corresponding to the PM peak period on
January 10, 2012 are shown in Tab.~\ref{cdc16-tab1}, where each entry is a
pair with the first element being the link index and the second
element being the scaled partial derivative value. For instance, the
entry $(0, 0.226)$ means 
\[
\frac{{\partial V\left(
		{{\bt^{{0}}},\bm}
		\right)}}{{\partial {t^{{0}}_a}}} = 0.226 \times
\max \left\{ {\frac{{\partial V\left(
			{{\bt^{{0}}},\bm}
			\right)}}{{\partial {t^{{0}}_a}}}; \,a = 0,
	\ldots ,23} \right\}
\] 
for link 0.
It is seen that the largest five values of ${\partial V\left(
	{{\bt^{{0}}},\bm} \right)}/{\partial
	{t^{{0}}_a}}$ correspond to links 19, 10, 8, 20, and
21 (red numbers on the left), and the largest five absolute values of ${\partial
	V\left( {{\bt^{{0}}},\bm}
	\right)}/{\partial {m_{a}}}$ correspond to links 23, 20, 19, 21, and
10 (blue numbers on the right). This suggests that, around the PM peak
period of January 10, 2012, the transportation management department
could have most efficiently reduced the total users' travel time by
taking actions with priorities on these links (e.g., improving road
conditions to reduce the free-flow travel time for links 19, 10, 8, 20,
and 21, and increasing the number of lanes to enlarge the flow capacity
for links 23, 20, 19, 21, and 10).

%
%
%
%
%
%



\begin{ack}
The authors would like to thank the Boston Region MPO, and Scott Peterson
in particular, for supplying the EMA data and providing us invaluable
clarifications throughout our work. 
\end{ack}

\bibliography{bib1}             

\end{document}